# Enumeration of saturated and unsaturated substituted N-heterocycles

Stefan Schuster* and Tatjana Malycheva

*Dept. of Bioinformatics, Matthias Schleiden Institute, Friedrich Schiller University of Jena, Ernst-Abbe-Pl. 2, 07743 Jena, Germany*

*Corresponding author, e-mail: stefan.schu@uni-jena.de

**Abstract**

Mathematical and computational approaches in chemistry and biochemistry fill a gap in respect to the analysis of the physicochemical features of compounds and their functionality and provide an overview of known as well as yet unknown, but hypothetically possible structures. Nitrogen-containing heterocycles such as aziridine, azetidine and pyrrolidine bear a high potential in pharmacology, biotechnology and synthetic biology. Here, we present a mathematical enumeration procedure for all possible azaheterocycles with at least one substituent depending on the number of atoms in the ring, in the sense of saturated and unsaturated congeners. One subgroup belonging to that substance class is constituted by ring-shaped amino acids with a secondary amino group, such as proline. A recursion formula is derived, which results in a modified Lucas number series. Moreover, an explicit formula for determining the number of such substances based on the Golden Ratio is given and a second one, based on binomial coefficients, is newly derived. This enumeration is a helpful tool for construction or complementation of virtual compound databases and for computer-assisted chemical synthesis route planning.

# Introduction

In view of many applications, it is instrumental to comprehend the combinatorial multitude of chemical substances of a given class. For example, in metabolomics, in the study of prebiotic or biotic evolution and in synthetic organic chemistry, it is important to know the maximum number of substances that can be expected. To enumerate theoretically possible structures, mathematical methods are helpful tools (Henze et al. 1931, Balaban et al. 1988, Haranczyk et al. 2010, Böcker et al. 2014, Fichtner et al. 2017). Enumeration is instrumental in constructing or complementing virtual compound databases and for computer-assisted chemical synthesis route planning. Combinatorial analysis, similar to enumeration studies on alcohols (Henze et al. 1931) and hydrocarbons (Balaban et al. 1988) was carried out on branched and unbranched aliphatic amino acids without any rings (Grützmann et al. 2011, Fichtner et al. 2017). There is a plethora of non-proteinogenic amino acids, occurring, for example, in toxic peptides (Arp et al. 2018, Kroteń et al. 2010, Zahradníčková et al. 2022). For instance, jessenipeptin involves

dehydro-2-aminobutyric acid, D-allo-threonine, L-diaminobutyric acid and D-proline (Arp et al. 2018). The latter belongs to the substance class under study here. Many other examples are reviewed in Fichtner et al. (2017). It can be expected that many more substances are synthesized in living organisms than have been discovered so far.

The aim of the present paper is to derive a method for calculating the number of nitrogen-containing heterocycles containing at least one substituent connected to a carbon in the ring by a single bond. We allow for varying numbers of double bonds in the ring. The enumeration is performed in dependence on ring size, while the substituent remains unchanged.

We focus on molecules in which the substituent is connected at a position adjacent to the nitrogen (2-position or $\alpha$-position). One subgroup belonging to that substance class is constituted by ring-shaped $\alpha$-amino acids with a secondary amino group, such as azetidine-2-carboxylic acid (see also next section), proline, pipecolic acid, and baikiain (Zahradníčková et al. 2022). The substituent is then a carboxyl group (COOH). However, the same enumeration procedure can be applied to azaheterocycles in which the substituent is located at another position as long as this does not lead to an additional plane of symmetry. Accordingly, we exclude the case where it is connected to the nitrogen.

Since the number of hydrogen atoms is not the same in all structures of a given size if the number of double bonds varies, our approach is about counting congeners rather than isomers. Methods for enumerating congeners are scarce so far (Haranczyk et al. 2010; Schuster et al. 2017).

Throughout this paper, we exclude adjacent (allenic) double bonds for two reasons. First, the bond angle would then amount to about 180°. That would be very difficult to realise in (closed) rings unless another two adjacent double bonds or a triple bond are situated on the opposite side of the ring. Second, allenic double bonds and triple bonds only occur very rarely in biomolecules (some occurrences are reviewed in Fichtner et al. 2017).

Several of the substances under consideration are rather unstable and, thus, short-lived. The question arises what the minimum lifetime of a molecule should be in order to be considered as a separate substance. Here, we use a very general approach in that we count separately, a priori, all conceivable structures of the type defined above with one exception: resonance structures (also known as Kekulé structures) (Lukovits et al. 2003, Hosoya 2005) of rings with fully conjugated double bonds. That is, resonance structures in a given ring are considered equivalent because the $\pi$ electrons are delocalised. However, for the sake of clarity in the mathematical presentation, we will first consider them separately and only later modify the calculation so as to combine them.

## Chemical and biochemical considerations

In many substances involving one or more (non-allenic) double bonds, cis-trans isomerism occurs. This means that the two single bonds adjacent to a double bond are situated on the same (cis) or on different sides (trans). Here we do not consider such isomers to be different. In fact, in three- to seven-membered rings, only the cis-isomer can occur for steric reasons (Neuenschwander et al. 2011). Moreover, we do not distinguish stereoisomers (such as L-proline and D-proline).

The smallest meaningful number for a ring is *n* = 3. Then, there are four possible ring structures, notably one involving single bonds only and three with one double bond each (Fig. 1). An example is provided by 2H-azirine-2-carboxylic acid, which is part of the antibacterial agent azirinomycin produced by *Streptomyces aureus* (Stapley et al. 1971).

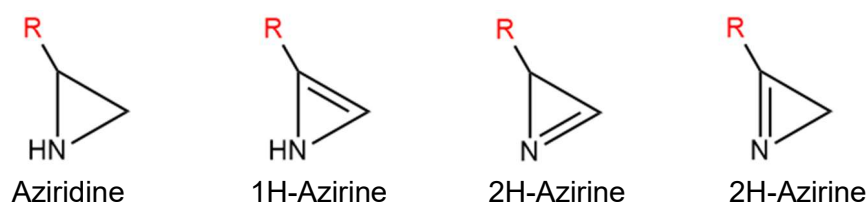

**Figure 1.** Three-membered N-heterocycles with one substituent R (-COOH; -CHO; -NH$_2$ etc.). Names refer to the unsubstituted rings. The left and right 2H-azirines are substituted at the 2 and 3 positions, respectively.

For *n* = 4, there are seven possible ring structures, notably one involving single bonds only, four with one double bond each and two with two non-adjacent double bonds each (Fig. 2). For the conjugated double bonds, two locations are feasible: in the representation shown in Fig. 2, they could be located in the vertical or horizontal direction. These resonance structures are equivalent and, thus, are finally counted only once. An example of a substance with a four-membered ring is azetidine-2-carboxylic acid, which is produced as a defence chemical against herbivory in several plants such as the lily of the valley (Fowden 1963).

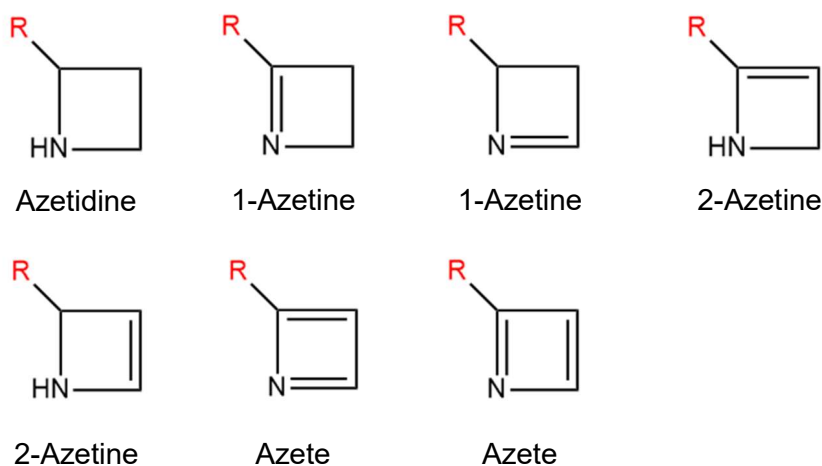

**Figure 2.** Four-membered N-heterocycles with one substituent R (-COOH; -CHO; -NH$_2$ etc.). Names refer to the unsubstituted rings. The two 1-azetine derivatives differ with respect to the position of the substituent, and so do the two 2-azetine derivatives. While both Kekulé structures of azete are shown, they are finally counted only once.

## Mathematical enumeration

*Counting resonance structures separately*

As mentioned above, it is convenient to first count resonance structures separately. We denote the numbers of structures by $y_n$. As mentioned above, the initial values read:

$$y_3 = 4, \quad y_4 = 7. \tag{1}$$

Now we derive a recursion formula for calculating the number series $y_k$. Assume we know all $y_k$ from $k = 3$ (see Fig. 1) up to $k = n$ and wish to calculate $y_{n+1}$. We consider the bond between the Cα and Cβ atoms and insert an additional carbon between these two atoms (see Fig. 3). In this way, we generate an $(n+1)$-membered ring.

If the original bond under consideration is a single bond, we make two single bonds out of it. If it is a double bond, we keep that type next to the Cα atom and make the second new bond a single bond. This is located between the new and original Cβ atoms, with the latter being the new Cγ atom.

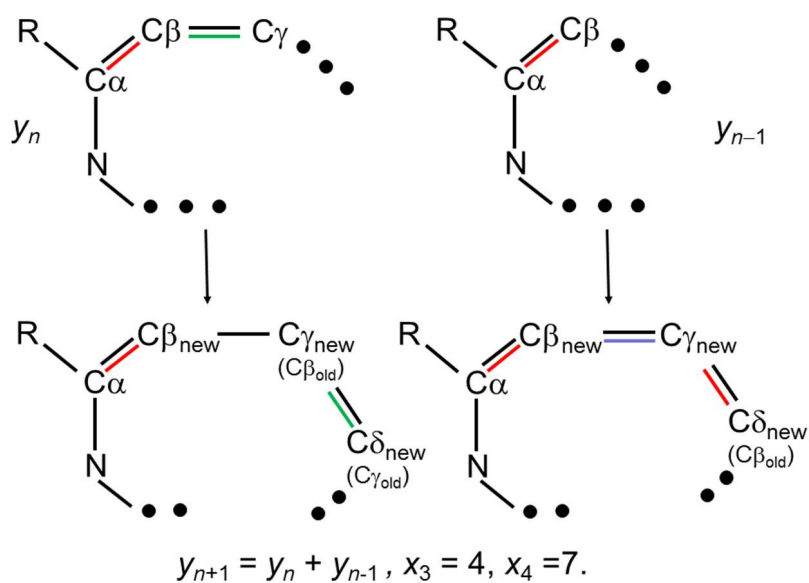

$$y_{n+1} = y_n + y_{n-1}, \quad x_3 = 4, \quad x_4 = 7.$$

**Figure 3.** Illustration of the recursive enumeration method. Solid dots, variable chain length. R, substituent. Red or green bar together with black bar: double bond that is maintained in the transition to ring size $n+1$. Blue bar together with black bar: double bond that is inserted in the extension provided that the bond between Cα and Cβ was a single bond. For further explanations, see text.

The question arises how we can add a double bond between these two carbon atoms. As we do not know the type of bond between the original Cβ and Cγ atoms (if $n > 3$) or between the original Cβ and N atoms (if $n = 3$) and exclude allenic bonds, we cannot simply add a double bond. Instead, we start from the $(n-1)$-membered ring and insert two additional carbons between the original Cα and Cβ atoms. We do that in such a way that we keep the type of bond next to the Cα atom. If it is a single bond, we will insert a double bond between the new

Cβ and Cγ and a single bond between the new Cγ and Cδ (or N). If it is a double bond, we will insert a single bond between the new Cβ and Cγ and a double bond between the new Cγ and Cδ (or N).

Combining the two procedures (starting at $n-1$ and $n$), we arrive at the recursion equation

$$y_{n+1} = y_n + y_{n-1} \qquad (2)$$

with the initial conditions (1).

There is no overlap between the molecules thus generated because the $y_n$ molecules generated by starting from length $n$ have a single bond between the new Cβ and Cγ, while the $y_{n-1}$ molecules generated by starting from length $n-1$ have a double bond at that position or, alternatively, the pattern double-single-double starting at the Cα atom. That pattern does not occur in the transition from $n$ to $n+1$ because it would imply two adjacent double bonds before the transition. Moreover, all possibilities of extending the molecules according to the defined rules are covered.

Eqs. (1) and (2) define the series of Lucas numbers (Koshy 2001, Taskara et al. 2010, Azarian 2012, Fowler et al. 2014). The series reads

$$y_n = 1, 3, 4, 7, 11, 18, 29, 47, 76, 123, 199, 322, 521, 843, \ldots \qquad (3)$$

where we included $y_1 = 1$ and $y_2 = 3$. It is named after French mathematician François Édouard Anatole Lucas. In the Online Encyclopedia of Integer Sequences (www.oeis.org), the Lucas series is, together with some of its properties, given under no. A000204. It is worth noting that the recursion formula (2) with the alternative initial conditions $x_1 = 1$, $x_2 = 1$ leads to the famous Fibonacci numbers (Koshy 2001).

There is a concept in graph theory that is related to the approach used here. A matching of a graph is a set of edges (arcs) without common vertices (Diestel 2000, Wagner and Wang 2023). That edge set corresponds to the double bonds in biomolecules because the latter must not be adjacent. The total number of matchings is the Hosoya index or topological index, which leads to the Fibonacci series in the case of linear graphs and to the Lucas series for monocyclic graphs (Hosoya 1973, 1988, Wagner and Wang 2023). The topological index has also been used in chemistry, notably for coding the complicated structures of cyclic and branched hydrocarbons and for predicting some of their thermodynamic quantities such as the boiling point (Hosoya 1973). Hosoya (1988, 2005) used that quantity to extend the famous Hückel rule.

The topological index is sometimes written as the sum of the numbers $p(G,k)$ of matchings including $k = 0, 1, 2$ etc. edges (with $G$ denoting the graph) (Koshy 2001). In the application under study here, these are the numbers of substances with 0, 1, 2 etc. double bonds. To our knowledge, we are the first to have noticed the relationship to double bonds and, thus, the use of the topological index for enumerating congeners with varying numbers of double bonds (Schuster et al. 2017, Fichtner et al. 2017).

As pointed out by Hosoya (1973), ring-shaped molecules can formally be considered even from $n=1$ on. A monogon and a digon may be defined, respectively, as a point graph and a graph with two points joined by two lines. We can formally define that there is only one structure of a one-membered ring. There are three matchings for the digon: no edge, one of the edges or the other one. While we started with the ring sizes of three and four and, accordingly with $y_3$ and $y_4$, we can go backwards in the recursion to $y_2 = 3$ and $y_1 = 1$.

Eq. (2) is a recursion equation in that the numbers for certain ring sizes are calculated from the numbers of smaller rings. It would be helpful to derive an explicit formula, by which the number $y_n$ can directly be calculated from the ring size, $n$. This can be achieved by an exponential ansatz $y_n = \lambda^n$. Substituting this into Eq. (2) leads to the quadratic equation $\lambda^2 - \lambda - 1 = 0$. The two solutions of that equation are then used as the bases of the linear combination of two exponential functions (Koshy 2001):

$$y_n = \left(\frac{1+\sqrt{5}}{2}\right)^n + \left(\frac{1-\sqrt{5}}{2}\right)^n \tag{4}$$

The coefficients in this linear combination are both equal to one; they have been determined from the initial values $y_1 = 1$, $y_2 = 3$. The quantity $(1 + \sqrt{5})/2$ is the famous Golden Ratio, 1.618... (Koshy 2001). As the absolute value of the second basis, $|1 - \sqrt{5}|/2$, is between zero and one, the influence of the second term in the explicit formula gets smaller and smaller as $n$ increases. Already from $n = 2$ on, we can simplify the formula to

$$y_n = \text{round}\left(\frac{1+\sqrt{5}}{2}\right)^n \tag{5}$$

This implies that the limit of the ratio of two consecutive numbers for large $n$ is the Golden Ratio. The analogous formula for the Fibonacci numbers reads, from $n = 0$ on (Koshy 2001):

$$F_n = \text{round}\left[\frac{1}{\sqrt{5}}\left(\frac{1+\sqrt{5}}{2}\right)^n\right] \tag{6}$$

*Alternative formula based on binomial coefficients*

Besides the explicit formula (6) for the Fibonacci numbers, there is an alternative formula in the form of a sum of binomial coefficients, which was derived by Lucas in 1876 (Koshy 2001, Azarian 2012). For the case of fatty acids and unbranched aliphatic amino acids, that formula has an instructive interpretation in terms of double bonds (Schuster et al. 2017): each binomial coefficient $\binom{n-k}{k}$ quantifies the number of structures involving exactly $k$ double bonds (corresponding to the $k$-combinations in mathematics). It involves $n-k$ rather than $n$ because double bonds must not be adjacent to each other.

Inspired by the reasoning in terms of binomial coefficients, one can derive an analogous formula for Lucas numbers:

$$y_n = \sum_{k=0}^{r}\binom{n+1-k}{k} - \sum_{k=0}^{s}\binom{n-3-k}{k} = F_{n+2} - F_{n-2} \tag{7}$$

where $r = \text{floor}[(n+1)/2]$ and $s = \text{floor}[(n-3)/2]$. The floor function denotes the largest integer that is less than or equal to its argument. Formula (7) also holds for $n = 1$ and $n = 2$ if we extend the Fibonacci series to negative indices: $F_{-1} = 1$ and $F_{-2} = -1$. The fact that each Lucas number can be written as the difference of two Fibonacci numbers has been known before (Koshy 2001, p. 97).

In terms of chemical structures, formula (7) can be explained as follows. We cut the ring at the N atom and thus transform it to an acyclic (straight) molecule. Formally, the N atom now occurs twice (Fig. 4). The first sum in Eq. (7) gives the number of congeners in straight molecules (Schuster et al. 2017). The formula there slightly differs because in fatty acids, no double bond is allowed next to the carboxyl group.

Now we have to take into account that in the cyclic molecule, there cannot be two adjacent double bonds next to the N atom. Therefore, we have to subtract all congeners that have double bonds both at the upper and lower ends of the straight molecule. Since at the second and second to last positions, only single bonds are allowed then, we need to subtract the possibilities in the central part, which involves four bond positions less (in the red ellipse in Fig. 4). Thus, the second sum in eq. (7) is running up to $n-3-k$ rather than $n+1-k$.

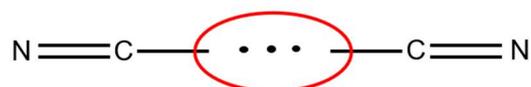

**Figure 4.** Schematic representation of the transformation of the ring molecule into a straight-chain aliphatic molecule by a cut at the heteroatom, to explain the derivation of formula (7). Alternatively, between the N atom (here shown twice) and any of the two adjacent C atoms, there might be a single bond. However, to explain the second term in Eq. (7), we here illustrate the case where there are double bonds at both positions.

*Combining resonance structures*

Now, we return to taking into account that rings with fully conjugated double bond systems have the property that the double bonds are delocalised. Thus, it cannot be said whether these bonds are located at odd-numbered (1, 3, ...) or even-numbered positions. This applies not only to aromatic rings (e.g. picolinic acid, which involves a six-membered ring) but also to non-aromatic rings with fully conjugated double bond systems (which do not satisfy the Hückel rule). Even if the latter rings subsist in non-planar conformations (e.g. boat conformation) to avoid anti-aromaticity, they undergo transitions to other conformations. Every now and then, such transitions lead to a transient planar conformation, which enables a delocalisation of electrons. Therefore, the numbers of congeners of all rings with even-numbered ring sizes are by one lower than given by the Lucas series.

Let $x_n$ denote the number of theoretically possible, ring-shaped congeners involving $n$ carbons, without distinguishing resonance structures. Formally including the numbers $x_1$ and $x_2$, we obtain the series:

$$x_n = 1, 2, 4, 6, 11, 17, 29, 46, 76, 122, 199, 321, 521, 842, ... \qquad (8)$$

Also this series is given in the Online Encyclopedia of Integer Sequences (www.oeis.org), notably under number A014217.

The monogon can only involve either the Cα atom or a nitrogen but not both. Thus, it does not correspond to a meaningful amino acid. In the digon, the Cα atom and nitrogen can only be linked by one bond, either a single or a double bond. In the context of amino acids, these two possibilities correspond to glycine and dehydroglycine (iminoacetic acid), which is an intermediate in the biosynthesis of thiamine (Jurgenson et al. 2009). Interestingly, the second number in series (8), which is 2, correctly describes the number of structures for *n* = 2. This is because the two hypothetical situations where the double bond is on one or the other side are equivalent.

Also the number series $x_n$ can be calculated explicitly, notably by a formula given in www.oeis.org (number A014217):

$$x_n = \text{floor}\{(\frac{1+\sqrt{5}}{2})^n\} \tag{9}$$

For example, $(1.618...)^4$ = 6.854..., which gives 6. To understand formula (9), one can start from eq. (4) and take into account that the basis of the second power term therein is −0.618..., which is the negative reciprocal of the Golden Ratio. The powers $(-0.618...)^n$ are numbers between −1 and 0 for odd *n* and between 0 and 1 for even *n*. Thus, $y_n$ is equal to floor$\{[(1 + \sqrt{5})/2]^n\}$ for odd *n* and floor$\{[(1 + \sqrt{5})/2]^n\}$ + 1 for even n. Together with the definition of $x_n$ in terms of $y_n$, this reasoning leads to eq. (9).

The alternative formula (7) can be modified as follows to give the number series (8):

$$x = \sum_{k=0}^{r} \binom{n+1-k}{k} - \sum_{k=0}^{s} \binom{n-3-k}{k} - \frac{1+(-1)^n}{2} \tag{10}$$

For every even *n*, the additional term equals −1, while for every odd *n*, it equals zero.

## Discussion

Here, we have presented a method for calculating the number of all possible azaheterocycles as a function of ring size, with the flexibility that various numbers of double bonds are allowed. We have excluded adjacent double bonds because they occur very rarely in ring-shaped molecules. We have given both a (relatively simple) recursion equation and two explicit formulas. One of these includes Fibonacci numbers. The resulting number series (www.oeis.org, accession no. number A014217) is closely related to the Lucas series known from number theory (Hosoya 1973, Koshy 2001, Taskara et al. 2010, Azarian 2012) in that every second number is equal to a Lucas number while the numbers in between are by one lower than in the Lucas series.

Enumeration by the same method is also applicable to cyclic azacycloalkane or azacycloalkene analogs with more than one substituent. An example is azirinomycin, which is 3-methyl-2H-azirine-2-carboxylic acid (Stapley et al. 1971). It is worth mentioning that rings without any substituent obey different enumeration rules because they show a higher symmetry. The same is true for some molecules with several identical substituents if an additional symmetry plane occurs.

Other applications of Lucas numbers in mathematical chemistry have been found earlier. In addition to the above-mentioned applications in chemistry (Hosoya 1973, 1988, 2005), that series is also useful to describe the numbers of Kekulé structures in certain classes of polycyclic aromatic compounds. For [*n*]circulenes, the number 2 has to be added to every second Lucas number (Bergan et al. 1987, Cyvin et al. 1991). The series itself can be used for a class of aromatic compounds for which Fowler et al. (2014) suggested the term lucacenes.

As mentioned in the Introduction, acyclic aliphatic amino acids have been enumerated earlier (Grützmann et al. 2011, Fichtner et al. 2017). For the unbranched substances including single and possibly double bonds, the famous Fibonacci series is obtained (Fichtner et al. 2017). Importantly, the ratio of two consecutive numbers tends to the Golden Ratio for the Fibonacci series, the Lucas series and for the number series (8) derived here.

Three- and four-membered heterocyclic rings are strained due to their small bond angles and thus, most of them are highly reactive and unstable (Behre et al. 2012, Banert et al. 2013). Moreover, 1H-azirines are much less stable than their tautomers 2H-azirines. This is because the nitrogen in the former rings contributes two electrons to the conjugated double bonds, so that the number of delocalized electrons equals four. As 1H-azirines are, in addition, planar, they are anti-aromatic and, thus, very unlikely to be found in natural products (Zavoruev et al. 1990, Banert et al. 2013). It is a difficult decision what the minimum lifetime of a molecule should be so that it is considered as a separate substance. We leave this decision to further studies and have here taken into account, a priori, all conceivable structures of the type under study with one exception: different resonance/Kekulé structures of aromatic compounds.

The term enumeration has two meanings. In the sense used in the present paper, it refers to calculating the number of possible structures. In contrast, constructive enumeration requires that all structures are listed successively in silico, for example, by a graph-theoretical approach (Lukovits et al. 2003). For small and moderate ring sizes, this can be done relatively easily, as we have done in Figs. 1 and 2. For larger rings, it will be of interest to establish a suitable code for systematic, constructive enumeration.

The presented method for enumeration bears manifold potential applications. In metabolomics, it is of interest to know how many different substances can be expected, and so it is in synthetic biology. In the latter field, this helps to assess how long it would take, approximately, to synthesize a complete library of them. Analogous considerations are relevant in patent applications. In the study of evolution, it is useful to analyse possible alternatives to the biomolecules found in extant living organisms. Implementation into virtual compound libraries, which are used in computer-assisted drug discovery, testing of hypotheses and experiment optimization, are potential promising applications (Faulon et al. 2005).

**Acknowledgment**: The authors thank Stephan Wagner (Uppsala), Severin Sasso (Leipzig) as well as Hans-Dieter Arndt and Rainer Beckert (Jena) for stimulating discussions. Financial support by the Deutsche Forschungsgemeinschaft (DFG, German Research Foundation) within the SFB 1127 ChemBioSys, grant no. 239748522, is gratefully acknowledged.